\documentclass[10pt,conference]{IEEEtran}
\IEEEoverridecommandlockouts
\usepackage{graphicx}

\usepackage{url}

\usepackage{color}

\usepackage{listings}

\usepackage[fleqn]{amsmath}

\usepackage{booktabs}

\usepackage{tablefootnote}

\usepackage{cite}

\usepackage{enumitem}

\definecolor{lightgray}{rgb}{.9,.9,.9}
\definecolor{darkgray}{rgb}{.4,.4,.4}
\definecolor{purple}{rgb}{0.65, 0.12, 0.82}

\lstdefinelanguage{JavaScript}{
  keywords={typeof, new, true, false, catch, function, return, null, catch, switch, var, if, in, while, do, else, case, break},
  keywordstyle=\color{blue}\bfseries,
  ndkeywords={class, export, boolean, throw, implements, import, this},
  ndkeywordstyle=\color{darkgray}\bfseries,
  identifierstyle=\color{black},
  sensitive=false,
  comment=[l]{//},
  morecomment=[s]{/*}{*/},
  commentstyle=\color{purple}\ttfamily,
  stringstyle=\color{red}\ttfamily,
  morestring=[b]',
  morestring=[b]"
}

\renewcommand\lstlistingname{Code.}

\makeatletter
\def\lst@makecaption{%
  \def\@captype{table}%
  \@makecaption
}
\makeatother

\makeatletter
\def\lst@lettertrue{\let\lst@ifletter\iffalse}
\makeatother
\newcommand{\IEEEauthorblockAOsaka}[1]{%
  \IEEEauthorblockA{%
    \IEEEauthorrefmark{1}%
    The University of Osaka, Japan, %
    #1@ist.osaka-u.ac.jp%
  }%
}
\newcommand{\IEEEauthorblockANDSU}[1]{%
  \IEEEauthorblockA{%
    \IEEEauthorrefmark{2}%
    Notre Dame Seishin University, Japan, %
    #1@m.ndsu.ac.jp%
  }%
}
\newcommand{\IEEEauthorblockAFukuchiyama}[1]{%
  \IEEEauthorblockA{%
    \IEEEauthorrefmark{3}%
    The University of Fukuchiyama, Japan, %
    #1@fukuchiyama.ac.jp%
  }%
}
\newcommand{\IEEEauthorblockANanzan}[1]{%
  \IEEEauthorblockA{%
    \IEEEauthorrefmark{4}%
    Nanzan University, Japan, %
    #1@nanzan-u.ac.jp%
  }%
}
\newcommand{\IEEEauthorblockAToshiba}[1]{%
  \IEEEauthorblockA{%
    \IEEEauthorrefmark{5}%
    Toshiba Corporation, Japan, %
    #1@toshiba.co.jp%
  }%
}

\begin{document}

\title{A Dataset of Software Bill of Materials for Evaluating SBOM Consumption Tools}

\author{
  \IEEEauthorblockN{%
    Rio~Kishimoto\IEEEauthorrefmark{1}, %
    Tetsuya~Kanda\IEEEauthorrefmark{2}, %
    Yuki~Manabe\IEEEauthorrefmark{3}, %
    Katsuro~Inoue\IEEEauthorrefmark{4}, %
    Shi~Qiu\IEEEauthorrefmark{5}, %
    Yoshiki~Higo\IEEEauthorrefmark{1}}
  \IEEEauthorblockAOsaka{\{r-kisimt, higo\}}
  \IEEEauthorblockANDSU{kanda}
  \IEEEauthorblockAFukuchiyama{manabe-yuki}
  \IEEEauthorblockANanzan{inoue599}
  \IEEEauthorblockAToshiba{shi1.qiu}
  \thanks{}
}
% \thanks{Supported by organization x.}

\maketitle

\begin{abstract}

  A Software Bill of Materials (SBOM) is becoming an essential tool for effective software dependency management.
  An SBOM is a list of components used in software, including details such as component names, versions, and licenses.
  Using SBOMs, developers can quickly identify software components and assess whether their software depends on vulnerable libraries.
  Numerous tools support software dependency management through SBOMs,
  which can be broadly categorized into two types: tools that generate SBOMs and tools that utilize SBOMs.
  A substantial collection of accurate SBOMs is required to evaluate tools that utilize SBOMs.
  However, there is no publicly available dataset specifically designed for this purpose,
  and research on SBOM consumption tools remains limited.
  In this paper, we present a dataset of SBOMs to address this gap.
  The dataset we constructed comprises 46 SBOMs generated from real-world Java projects,
  with plans to expand it to include a broader range of projects across various programming languages.
  Accurate and well-structured SBOMs enable researchers to evaluate the functionality of SBOM consumption tools and identify potential issues.
  We collected 3,271 Java projects from GitHub and generated SBOMs for 798 of them using Maven with an open-source SBOM generation tool.
  These SBOMs were refined through both automatic and manual corrections to ensure accuracy,
  currently resulting in 46 SBOMs that comply with the SPDX Lite profile, which defines minimal requirements tailored to practical workflows in industries.
  This process also revealed issues with the SBOM generation tools themselves.
  The dataset is publicly available on Zenodo (DOI: 10.5281/zenodo.14233414).

\end{abstract}


\section{Introduction}
\label{sec:introduction}

Managing software dependencies is an increasingly important task due to its growing complexity.
Inadequate software dependency management has resulted in delayed responses to vulnerabilities hidden within dependent libraries~\cite{saner2021-python-package-vulns-analysis,icsme2020-library,kula_developers_2018}.
As a promising tool for addressing these challenges, a Software Bill of Materials (SBOM) has gained attention~\cite{whitehouse-executive-order,eu-cra}.
An SBOM is a comprehensive inventory of components used in software,
including details such as component names, versions, and licenses.
Using SBOMs, developers can quickly identify the components their software depends on
and assess whether these dependencies include vulnerable libraries.

Numerous tools currently support software dependency management through SBOMs,
which can be broadly categorized into two types: tools that generate SBOMs and tools that consume (utilize) SBOMs.
SBOM generation tools analyze software source code, extract information about its components, and produce an SBOM.
SBOM consumption tools, on the other hand, analyze an SBOM to provide insights.
For example, grype~\cite{grype} is an SBOM consumption tool that accepts SBOMs as input and identifies vulnerabilities in the software components described within.

While the performance and limitations of SBOM generation tools have been studied extensively~\cite{linuxfoundation-sbom-cybersecurity-readiness,ieeesp-challenges-of-producing-sbom-for-java,ares24-accuracy-evaluation-of-sbom-tools,icse2023-empirical-study-on-sbom,icse2024-boms-away},
research on SBOM consumption tools remains limited.
One of the reasons is the lack of a publicly available dataset specifically designed for evaluating SBOM consumption tools.
To effectively evaluate SBOM consumption tools, a substantial collection of accurate and well-structured SBOMs is necessary.
The examples of SBOM~\cite{spdx-examples,bom-examples} typically describe software with few dependencies, limiting their utility for evaluating tools.
Chainguard also provides a collection of SBOMs~\cite{bom-shelter}, but these are often automatically generated or obtained from the Internet,
which means their accuracy cannot be guaranteed.

To address this gap, we present a dataset of SBOMs specifically designed to evaluate SBOM consumption tools,
with a focus on two major SBOM applications: vulnerability management and license management.
We collected 3,271 Java projects from GitHub and generated SBOMs for 798 of them using Maven with an open-source SBOM generation tool.
These SBOMs were refined through both automatic and manual corrections to ensure accuracy,
currently resulting in 46 SBOMs that comply with the SPDX Lite profile~\cite{spdx-lite-v2.2.2}, which defines minimal requirements tailored to practical workflows in industries.
These accurate and well-structured SBOMs enable researchers to thoroughly evaluate the functionality of SBOM consumption tools and identify potential issues.

\section{Dataset Requirements and SBOM Format}
\label{sec:requirements}

There are two major SBOM formats: SPDX~\cite{spdx-about} and CycloneDX~\cite{cyclonedx}.
\lstlistingname~\ref{code:spdx-example} shows an excerpt of an SBOM written in the SPDX format.
This example describes information about a widely used Java library, log4j-core version 2.10.0.
For instance, the library name (name field, line 1), its version (versionInfo field, line 8), and its license (licenseConcluded field, line 6) are recorded in the corresponding fields.
SPDX defines numerous additional fields to represent various types of software-related information,
and CycloneDX offers a similar set of fields.
Since most of these fields are optional, the choice of which fields to include in an SBOM depends on its intended use case.

The primary use case of our dataset is the evaluation of SBOM consumption tools,
with a focus on two major SBOM applications: vulnerability management and license management.
Therefore, it is sufficient for the SBOMs in our dataset to include the essential information required for these two use cases.

In this paper, we adopt the SPDX format because it provides a set of mandatory fields called the SPDX Lite profile\cite{spdx-lite-v2.2.2}.
This profile defines a minimal set of fields tailored to actual workflows in industries.
By complying with this profile, we ensure that the SBOMs in our dataset contain the critical information needed for the major use cases while maintaining simplicity and relevance.

\begin{figure}
  \begin{lstlisting}[label={code:spdx-example},caption={Excerpt of an SBOM in SPDX Format for a Java Project}]
{
  "name": "log4j-core",
  "SPDXID": "SPDXRef-Package-log4net",
  "downloadLocation": "https://repo1.maven.org/maven2/org/apache/logging/log4j/log4j-core/2.10.0/log4j-core-2.10.0.jar",
  "filesAnalyzed": false,
  "licenseConcluded": "Apache-2.0",
  "copyrightText": "NOASSERTION",
  "versionInfo": "2.10.0",
  "externalRefs": [{
    "referenceCategory": "PACKAGE-MANAGER",
    "referenceLocator": "pkg:maven/org.apache.logging.log4j/log4j-core@2.10.0",
    "referenceType": "purl"
  }],
  "supplier": "Organization: The Apache Software Foundation",
  "homepage": "https://logging.apache.org/ ... "
}
\end{lstlisting}
\end{figure}

\section{Methodology}
\label{sec:methodology}

\begin{figure*}[tb]
  \centering
  \includegraphics[width=0.6\textwidth]{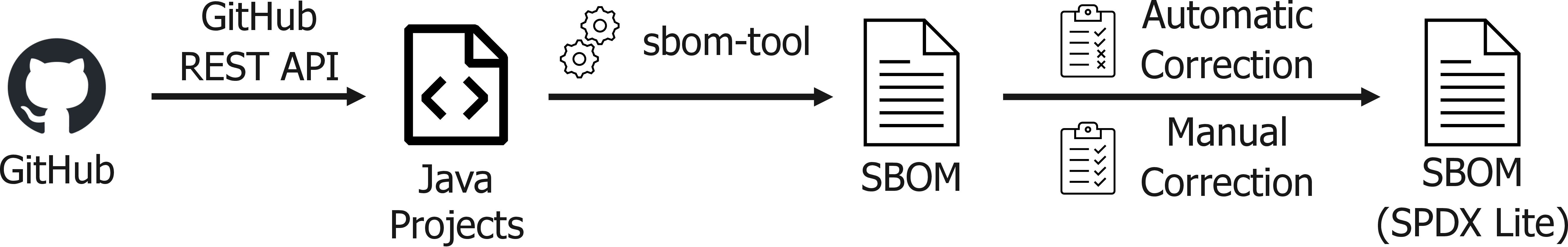}
  \caption{Overview of the dataset construction process}
  \label{fig:overview}
\end{figure*}

Figure \ref{fig:overview} shows the overview of the dataset construction process.
As the target programming language, we selected Java because it is widely used in various software projects and many projects are available on GitHub.
The dataset construction process is divided into four steps: Java project collection, SBOM production with sbom-tool, automatic correction, and manual correction.
First, we search Java projects on GitHub and clone their repositories.
Next, we analyze the source code of the projects with an open-source SBOM generation tool: sbom-tool\cite{sbom-tool},
and create initial SBOMs.
After getting initial SBOMs, we correct missing information in the following steps.
We automatically correct the SBOMs to fill in the essential fields with the information in the configuration files for Maven (pom.xml).
Finally, we manually correct the SBOMs and create the SBOMs that follow the SPDX Lite profile.

\subsection{Java Project Collection}

We collected Java projects from GitHub.
We obtained an initial list of Java projects using the GitHub API,
by searching for repositories that met the following conditions:
(1) whose primary language is Java;
(2) that have at least 1000 stars;
(3) the repository is not archived.
These conditions were chosen to ensure that the projects are active and popular.

We obtained 3,271 GitHub repositories matching these conditions.
We cloned these projects to analyze the source code of the projects in the next SBOM production step.
3,254 projects were successfully cloned and 17 projects failed due to some errors such as the path length limitation on the Windows environment.

\subsection{SBOM Production with sbom-tool}

We produced initial SBOMs for the cloned projects with sbom-tool\cite{sbom-tool}, an open-source SBOM generation tool.
This popular and actively maintained tool supports the generation of SBOMs for Java projects using Maven as a build tool.

First, we selected Java projects that use Maven as a build tool from the cloned projects.
This is because sbom-tool does not support SBOM generation for Java projects using other build tools than Maven.
We consider that a project uses Maven if it contains at least one pom.xml file, a configuration file for Maven.
We gained 798 projects using Maven.

For these projects, we created SBOMs with sbom-tool.
We used version 2.9.9 of sbom-tool, and ran it with the following command:

% \lstinline[basicstyle=\small]|sbom-tool generate -b <output-dir> -bc <src-dir> -pn <name> -pv <version> -ps <supplier>|
sbom-tool generate -b \textit{output-dir} -bc \textit{src-dir} -pn \textit{name} -pv \textit{version} -ps \textit{supplier}

\begin{description}[labelwidth=4.5em,labelindent=0.5em]
  \item[\rm \textit{output-dir}] directory path for the output SBOM.
  \item[\rm \textit{src-dir   }] project root directory path.
  \item[\rm \textit{name      }] project name (value: repository name).
  \item[\rm \textit{version   }] project version (value: latest commit hash).
  \item[\rm \textit{supplier  }] project supplier (value: repository owner name).
\end{description}

The primary language used in the collected projects is Java, but some projects contain code written in other languages.
It means that they may have dependencies that are not managed by Maven.
Sbom-tool can detect a package manager used in a project, so we checked the log output and excluded a project from the SBOM production candidates if sbom-tool detects package managers other than Maven.

\subsection{Automatic Correction}

SBOMs produced by sbom-tool do not contain values of some fields.
Table \ref{table:essential-fields-dependency} shows the essential fields about a package (e.g. library) in the SPDX Lite profile.
The first column shows the essential SPDX fields, and the next two columns indicate whether information can be
retrieved with SBOM production with sbom-tool and automatic correction steps respectively.
In this table, ``+'' means that the field is filled in for all projects,
while ``o'' means that the field is filled in for some projects because the data source is optional.
As the column labeled ``sbom-tool'' shows, only the following fields are filled with sbom-tool:
Package Name, Package SPDX Identifier, Package Version, Files Analyzed, and External Reference.
Therefore, we need to fill in the other fields to create SBOMs that comply with the SPDX Lite profile.

Most of the Java components are published on the Maven Central Repository\footnote{https://central.sonatype.com/}.
It provides a REST API\footnote{https://central.sonatype.org/search/rest-api-guide/}, and it enables us to obtain the information to fill some of the remaining fields.
Table \ref{table:automatic-correction-source} shows the correspondence between fields and information that can be obtained using the API.
We can get the file name of the jar file of a component (for the Package File Name field),
and the URL to its jar file (for the Package Download Location field) through the API.
The API also provides the URL to the configuration file of the component (pom.xml), which contains some additional information about the component.
Using the information available on the Maven Central Repository, we can populate six fields for a component:
\textit{Package File Name, Package Supplier, Package Download Location, Package Home Page, Concluded License, and Declared License}.
We wrote a script to obtain information from the API and fill in the fields.
In this step, we automatically corrected the SBOMs by running the script.

In the Maven ecosystem, component names are case-sensitive. However, sbom-tool does not handle case sensitivity correctly, which occasionally causes the script to fail when attempting to fetch information, even if the component is available on the Maven Central Repository.
Additionally, the script cannot retrieve information for components not published on the Maven Central Repository.
To address these limitations, we excluded projects with such dependencies from the candidates for SBOM production.
As a result, we successfully generated 46 SBOMs during the automatic correction step.

\begin{table}[tb]
  \centering
  \caption{Essential fields about Package}
  \label{table:essential-fields-dependency}
  \begin{tabular}{lll}
    \toprule
    SPDX Field name           & sbom-tool & Automatic  \\
                              &           & Correction \\
    \midrule
    Package Name              & +         &            \\
    %\midrule
    Package SPDX Identifier   & +         &            \\
    %\midrule
    Package Version           & +         &            \\
    %\midrule
    Package File Name         &           & +          \\
    %\midrule
    Package Supplier          &           & o          \\
    %\midrule
    Package Download Location &           & +          \\
    %\midrule
    Files Analyzed            & +         &            \\
    %\midrule
    Package Home Page         &           & o          \\
    %\midrule
    Concluded License         &           & o          \\
    %\midrule
    Declared License          &           & o          \\
    %\midrule
    Comments on License       &           &            \\
    %\midrule
    Copyright Text            &           &            \\
    %\midrule
    Package Comment           &           &            \\
    %\midrule
    External Reference        & +         &            \\
    \bottomrule
  \end{tabular}
\end{table}

\begin{table}[tb]
  \centering
  \caption{Information available from Maven Central for automatic correction}
  \label{table:automatic-correction-source}
  \begin{tabular}{ll}
    \toprule
    SPDX Field name           & Information on Maven Central         \\
    \midrule
    Package File Name         & Name of the jar file                 \\
    %\midrule
    Package Supplier          & Organization or Developers (pom.xml) \\
    %\midrule
    Package Download Location & URL of the jar file                  \\
    %\midrule
    Package Home Page         & URL of the homepage (pom.xml)        \\
    %\midrule
    Concluded License         & Licenses (pom.xml)                   \\
    %\midrule
    Declared License          & Licenses (pom.xml)                   \\
    \bottomrule
  \end{tabular}
\end{table}

\subsubsection{Package File Name \& Package Download Location}

The API provides the name of the jar file of a component and the URL to its jar file.
We can use these two pieces of information as the values of the Package File Name field and the Package Download Location field, respectively.

\subsubsection{Package Supplier}

A developer can \textit{optionally} include the organization to which a project belongs or the developer information of the component in the pom.xml.
If the organization information is included in the pom.xml, we use it as the value of the Package Supplier field;
otherwise, if a developer is included instead, the Package Supplier field is populated with that information.
If multiple developers are included, we cannot automatically determine which developer to use as the value of the Package Supplier field,
so the field is left blank.

\subsubsection{Package Home Page}

A developer can \textit{optionally} include the URL to the project's homepage in pom.xml.
%This information is optional.
If the information is included in pom.xml, we use it as the value of the Package Home Page field.

\subsubsection{Concluded License \& Declared License}

A developer can \textit{optionally} include the license information of the project itself in pom.xml.
%This information is also optional.
The license of the project is described using the name of the license and the URL of the license in pom.xml.
In SPDX format, well-known licenses are represented by their SPDX License Identifiers (e.g. Apache License Version 2.0 is represented by Apache-2.0).
If the name of the license or the URL of the license in pom.xml exactly matches that of a well-known license,
use its SPDX License Identifier as the value of the Concluded License field and the Declared License field.
There are some variations in the spelling of license names and URLs,
the license information in pom.xml does not often match the well-known licenses exactly.
In these cases, we use the information in pom.xml as is for the value of the two fields, and manually check and fix them in the next Manual Correction step.

\subsection{Manual Correction}

After the automatic correction step, there are still some fields whose values are empty or incorrect.
During the manual correction step, we reviewed and corrected these fields to ensure that the SBOMs complied with the SPDX Lite profile.
Across the 46 SBOMs generated in the automatic correction step, a total of 1,752 fields were manually corrected.

To fill the fields, we manually collected information from the following sources:
\begin{enumerate}
  \item pom.xml of the component
  \item Web pages whose URL is written in the pom.xml
  \item Web pages found by searching on the Internet
\end{enumerate}

First, we checked the information written in pom.xml of the component.
In the automatic correction step, our script sometimes failed to parse pom.xml correctly.
We checked these pom.xml manually and filled the fields of the SBOMs.

Second, we utilized information from the webpages whose URL is listed in pom.xml.
pom.xml contains the URL of the project's homepage, the URL of the issue tracker, and the URL of the source code repository (e.g. SourceForge, GitHub).
We attempted to obtain the missing information from those web pages.
If a URL is already broken, we checked the content of the web page using the Wayback Machine\footnote{http://web.archive.org/}.

In most cases, we can fill in the fields using the information collected in the above two steps.
However, in some cases, we could not find the information we needed from the pom.xml or the web pages.
In such cases, we searched for the source code repository of the component on the Internet.
If we found the source code repository, we extracted some information from it.
% 値が定められなかった場合はNOASSERTIONを入れる（今回のフィールドで空文字列を入れることになるフィールドは無い）
If we could not find the information we needed from the above sources to fill a field, we filled it with ``NOASSERTION'', which is a special value defined in the SPDX specification to indicate that the information is not available.

After correcting all fields, we validated the SBOMs using the SPDX Online Tool\footnote{https://tools.spdx.org/app/validate/}, an official tool provided by SPDX, to ensure that they comply with the SPDX format.

\section{Dataset Structure}
\label{sec:dataset}

The structure of the dataset is shown in Table \ref{table:dataset-structure}.
The dataset contains repositories.json, which is a list of Java projects.
The list includes the project ID, the name of the project, the project URL, the Git commit hash used to generate the SBOM, the number of dependencies of the project, and the number of stars on GitHub. % (\lstlistingname~\ref{code:repositories}).
For each project, we created a directory named ``[ID]\_[Name (/ is replaced with +)]''.
It contains an SBOM file (sbom.spdx.json) and a file describing the data sources of the SBOM (sbom.data-sources.json).

The users of the dataset can select SBOM files based on the information in repositories.json and utilize them as input for SBOM consumption tools to evaluate their functionality.
Additionally, detailed information about the data sources used to generate an SBOM is available in sbom.data-sources.json.
The file includes the type of the source for each field value (auto or manual), the name of the source, and any associated URL.
These details help users understand the process of SBOM creation and ensure transparency within the dataset.

\begin{table}[tb]
  \centering
  \caption{Structure of the dataset}
  \label{table:dataset-structure}
  \begin{tabular}{lp{4.4cm}}
    \toprule
    path                                           & description                                      \\
    \midrule
    /repositories.json                             & The list of Java projects in the dataset         \\
    %\midrule
    /[id]\_[name]/                                 & The directory for each Java project              \\
    %\midrule
    % 毎回ディレクトリ名を書くと長くて，2列ぶち抜きの表にすると別ページ送りにされるので，インデントと縦棒+ハイフンでディレクトリ内にあることを示す
    \hspace{1em} \textbar-- sbom.spdx.json         & The SBOM file for the project                    \\
    %\midrule
    \hspace{1em} \textbar-- sbom.data-sources.json & The file describing the data sources of the SBOM \\
    \bottomrule
  \end{tabular}
\end{table}

% \begin{figure}[tb]
%   \begin{lstlisting}[label={code:repositories},caption={Excerpt of repositories.json}]
% {
%   "FullName": "mybatis/mybatis-3",
%   "CloneUrl": "https://github.com/mybatis/mybatis-3.git",
%   "StargazersCount": 19721,
%   "UpdatedAt": "2024-10-03T03:58:30+00:00",
%   "Id": 8205602
% },
% \end{lstlisting}
% \end{figure}

\section{Limitations}
\label{sec:limitations}

% - Mavenのpomファイルに書かれている内容が誤っていることがあった
%  - 自動で取得した情報が誤っている可能性がある
% - 数が少ないかも
%  - プロジェクトは複雑な依存関係を持っているので，ある程度使えるはず
%    （どのぐらいの依存ライブラリの情報を各SBOMが含んでいるかの話はどこかでした方が良いかも）
%  - 今後増やす予定
% - 手作業なので途中で誤りが起こっているかも（data sourceの情報を提供しているから言及しなくても許されるかも）

In the automatic correction step, we relied on information obtained from the component's pom.xml file to populate certain fields.
However, during manual correction, we found that the pom.xml file occasionally contains inaccurate information.
For example, if a project is a fork of another and its pom.xml file has not been updated, it may include incorrect details.
Consequently, some values populated during the automatic correction step may contain errors.
% TODO: もうちょっと具体的に例を書きたいけど，どのプロジェクトでこれが起こっていたかメモし忘れていたので，間に合えば追記する
% （GitHub上でForkしてくれていれば，Fork元のpom.xmlと比較して検出が可能かもしれないけど，ソース管理システムへのリンクだけ更新してくれてるとは思えないので……）

Currently, the dataset comprises 46 SBOMs, which may be insufficient for a comprehensive evaluation of SBOM consumption tools.
Nonetheless, the projects included in the dataset have complex dependency relationships, making it valuable for certain evaluation scenarios.
In future work, we aim to expand the dataset by incorporating additional SBOMs for Java projects
and extending it to SBOMs for projects written in other programming languages.

\section{Conclusion}
\label{sec:conclusion}

This paper presents a dataset of SBOMs for evaluating SBOM consumption tools.
It comprises 46 SBOMs generated from real-world Java projects collected from GitHub.
The SBOMs were created using an open-source SBOM generation tool and refined through manual corrections to ensure accuracy.
This dataset has room for improvement in terms of both its quality and quantity,
so we plan to refine the data and expand it to include a broader range of projects across various programming languages.

Through the dataset construction process, we identified several issues with the SBOM generation tools.
Sbom-tool does not properly handle case sensitivity in component names, potentially leading to false negatives in vulnerability detection by SBOM consumption tools.
Additionally, creating an SBOM with all the necessary information currently requires significant manual effort.
To facilitate effective dependency management using SBOMs, it is crucial to address these limitations and minimize the manual work needed to produce accurate and comprehensive SBOMs.

\subsection*{Acknowledgements}

This work was supported by JSPS KAKENHI Grant Numbers
% 肥後先生
JP24H00692, JP21K18302, JP21H04877, JP23K24823, JP22K11985,
% 神田先生
JP24K14895,
% 眞鍋先生
JP21K02862,
% 井上先生
JP23K28065, and Nanzan University Pache Research Subsidy I-A-2 for the 2024 academic year.

% Referenceは別ページ
\clearpage

\bibliographystyle{IEEEtranS}
\bibliography{reference,evaluation-source}

\end{document}